\documentclass[english]{iopart}

\usepackage{graphicx}
\newcounter{fig}
\begin{document}

\title[High order Fuchsian ODE]{\Large
High order Fuchsian equations for the square lattice
Ising model: $\chi^{(6)}$  }

\author{ 
S. Boukraa$^\dag$, S. Hassani$^\S$, I. Jensen$^\ddag$,
J.-M. Maillard$^{||}$ and N. Zenine$^\S$}
\address{\dag LPTHIRM and D\'epartement d'A{\'e}ronautique,
 Universit\'e de Blida, Algeria}
\address{\S  Centre de Recherche Nucl\'eaire d'Alger, 
2 Bd. Frantz Fanon, BP 399, 16000 Alger, Algeria}
\address{$\ddag$ ARC Centre of Excellence for
Mathematics and Statistics of Complex Systems 
Department of Mathematics and Statistics,
The University of Melbourne, Victoria 3010, Australia}
\address{$||$ LPTMC, Universit\'e de Paris 6, Tour 24,
 4\`eme \'etage, case 121, 
 4 Place Jussieu, 75252 Paris Cedex 05, France} 
\ead{maillard@lptmc.jussieu.fr,   
 I.Jensen@ms.unimelb.edu.au, njzenine@yahoo.com, boukraa@mail.univ-blida.dz}

\begin{abstract}
This paper deals with $\tilde{\chi}^{(6)}$, the six-particle contribution
to the magnetic susceptibility of the square lattice Ising model.
We have generated, modulo a prime, series coefficients for $\tilde{\chi}^{(6)}$.
The length of the series is sufficient to produce the corresponding
Fuchsian linear differential equation (modulo a prime).
We obtain the Fuchsian linear differential equation that annihilates the ``depleted" series
$\Phi^{(6)}\,=\,\, \tilde{\chi}^{(6)}\, 
- {2 \over 3} \tilde{\chi}^{(4)}\, + {2 \over 45} \tilde{\chi}^{(2)}$.
The factorization of the corresponding differential operator is performed
using a method of factorization modulo a prime introduced in a previous paper.
The ``depleted" differential operator is shown to have a structure similar to the 
corresponding operator for $\tilde{\chi}^{(5)}$.
It splits into factors of smaller orders, with the left-most factor of order six
being equivalent to the symmetric fifth power of the linear differential 
operator corresponding to the elliptic integral $E$.
The right-most factor has a direct sum structure, and
using series calculated modulo several primes, all the factors in the direct sum
have been reconstructed in exact arithmetics.
\end{abstract} 

\vskip .5cm


\noindent {\bf AMS Classification scheme numbers}: 34M55, 
47E05, 81Qxx, 32G34, 34Lxx, 34Mxx, 14Kxx 


%\tableofcontents

\section{Introduction and recalls}
\label{intro}

The magnetic susceptibility (high temperature $\chi_{+}$  and
low temperature $\chi_{-}$)  of the square lattice Ising model is
given by~\cite{wu-mc-tr-ba-76}
\begin{eqnarray}
\label{plus}
 \chi_{+}(w) \, \, =  \, \, \, \sum \chi^{(2n+1)}(w) \,\, = \, \,\,
{{1} \over {s}} \cdot (1 - s^4)^{\frac{1}{4}} \cdot \sum \tilde{\chi}^{(2n+1)}(w)
\end{eqnarray}
and 
\begin{eqnarray}
\label{chimoins}
 \chi_{-}(w)\, \,  = \,\,\, \sum \chi^{(2n)}(w) \,\, = 
 \, \,\, (1 - 1/s^4)^{\frac{1}{4}} \cdot \sum \tilde{\chi}^{(2n)}(w).
\end{eqnarray}
in terms of the self-dual temperature variable $w=\frac{1}{2}s/(1+s^2)$,
with $s =\,\sinh(2J/kT)$.
The $n$-particle contributions $\tilde{\chi}^{(n)}$ are given by $n-1$ dimensional
integrals~\cite{nickel-99,nickel-00,pal-tra-81,yamada-84},
\begin{eqnarray}
\label{chi3tild}
\tilde{\chi}^{(n)}(w)\,\,=\,\,\,\, {\frac{1}{n!}}  \cdot 
\Bigl( \prod_{j=1}^{n-1}\int_0^{2\pi} {\frac{d\phi_j}{2\pi}} \Bigr)  
\Bigl( \prod_{j=1}^{n} y_j \Bigr)  \cdot   R^{(n)} \cdot
\,\, \Bigl( G^{(n)} \Bigr)^2, 
\end{eqnarray}
where\footnote[3]{The Fermionic term $\,G^{(n)}$ has several 
representations~\cite{nickel-00}.} 
\begin{eqnarray}
\label{Gn}
G^{(n)}\,=\,\, \prod_{1\; \le\; i\;<\;j\;\le \;n} \, h_{ij}, \,  \quad 
h_{ij}\,=\,\,
{\frac{2\sin{((\phi_i-\phi_j)/2) \cdot \sqrt{x_i \, x_j}}}{1-x_ix_j}}, 
\end{eqnarray}
and
\begin{eqnarray}
\label{Rn}
R^{(n)} \, = \,\,\, \,  {\frac{1\,+\prod_{i=1}^{n}\, 
x_i}{1\,-\prod_{i=1}^{n}\, x_i}}, 
\end{eqnarray}
with
\begin{eqnarray}
\label{thex}
&&x_{i}\, =\,\,\,  \,  \frac{2w}{1-2w\cos (\phi _{i})\, 
+\sqrt{\left( 1-2w\cos (\phi_{i})\right)^{2}-4w^{2}}},  
\\
\label{they}
&&y_{i} \, = \, \, \,
\frac{2w}{\sqrt{\left(1\, -2 w\cos (\phi _{i})\right)^{2}\, -4w^{2}}}, 
\quad \quad  \quad  \quad \sum_{j=1}^n \phi_j=\, 0  
\end{eqnarray}

As $n$ grows the series generation in the variable $w$ of the integrals
(\ref{chi3tild}) becomes very time consuming. In~\cite{bo-ha-ma-ze-07b}
calculations modulo a prime were performed on simplified integrals  $\Phi_H^{(n)}$
and this work demonstrated that most of the pertinent information 
(singularities, critical exponents, ...) can be obtained from linear ODEs known 
modulo a prime corresponding to the integrals $\Phi_H^{(n)}$. In order to go beyond 
$\tilde{\chi}^{(4)}$ this strategy was adopted previously for the  5-particle 
contribution $\tilde{\chi}^{(5)}$  \cite{bo-gu-ha-je-ma-ni-ze-08,bo-bo-gu-ha-je-ma-ze-09}
and here for the 6-particle contribution $\tilde{\chi}^{6)}$ .

In a previous paper~\cite{bo-gu-ha-je-ma-ni-ze-08}  
 massive computer calculations were performed on
$\tilde{\chi}^{(5)}$, $\tilde{\chi}^{(6)}$ and $\chi$ (in exact arithmetics
and/or modulo a prime).
These calculations confirmed previously conjectured singularities for
the linear ODEs of the  $ \, \tilde{\chi}^{(n)}$'s 
as well as their critical exponents, and shed some light
on important physical problems such as the existence of a natural boundary for 
 the susceptibility of the square Ising model and the subtle resummation
of logarithmic behaviours of the $n$-particle contributions $ \, \tilde{\chi}^{(n)}$ 
to give rise to the power laws of the full susceptibility $\chi$.
As far as $\, \chi^{(5)}$ is concerned, the linear ODE for 
 $ \, \tilde{\chi}^{(5)}$ was   
found modulo a single  prime~\cite{bo-gu-ha-je-ma-ni-ze-08}
and it is of minimal order 33.

\subsection{Results on $\tilde{\chi}^{(5)}$}

In~\cite{bo-bo-gu-ha-je-ma-ze-09} the linear differential
operator for $ \, \tilde{\chi}^{(5)}$ was carefully analysed. In particular 
it was found that the minimal order linear differential operator 
for $ \, \tilde{\chi}^{(5)}$  can be reduced 
to a  minimal order linear differential operator $L_{29}$ of order 29 
for the linear combination
\begin{eqnarray}
\label{thePhi5}
\Phi^{(5)} \,=\,\, \, \tilde{\chi}^{(5)}\, -{1 \over 2}\, \tilde{\chi}^{(3)}
 \, + {\frac{1}{120}}\, \tilde{\chi}^{(1)}.
\end{eqnarray}
We shall use the term ``depleted" series for a series obtained by substracting from
$ \, \tilde{\chi}^{(n)}$ a definite amount of the lower $n$-particle
contributions $ \, \tilde{\chi}^{(n-2k)}$, $k=1, 2, \cdots$, as in (\ref{thePhi5}), such that
the differential operator annihilating the depleted series is of lower order.
Since the depleted series is annihilated by an ODE of lower order, it follows
that in the ODE for the original series, we must
 have the occurrence of a direct sum structure.
It was found~\cite{bo-bo-gu-ha-je-ma-ze-09} that the linear differential
operator $L_{29}$, can be factorised as a product of an order five, an order
twelve, an order one, and an order eleven linear differential operator 
  \begin{eqnarray}
\label{63}
L_{29} \,=\,\, \,\, L_5 \cdot L_{12} \cdot \tilde{L}_1 \cdot L_{11},
\end{eqnarray}
where the order eleven linear differential operator 
 has a direct-sum decomposition
\begin{eqnarray}
\label{direct}
L_{11} \,=\,\,\, 
 \left( Z_2 \cdot N_1 \right) \oplus V_2 \oplus 
\left( F_3 \cdot F_2 \cdot L_1^s \right).
\end{eqnarray}
$\, Z_2$ is a second order operator also occurring
in the factorization of the linear differential operator~\cite{ze-bo-ha-ma-04}
associated with $ \, \tilde{\chi}^{(3)}$ and it corresponds to a 
{\em modular form of weight one}~\cite{bo-bo-ha-ma-we-ze-09}.
$\, V_2$ is a second order operator equivalent to
the second order operator  associated with   $ \, \tilde{\chi}^{(2)}$ (or
equivalently to the complete elliptic integral $E$).
 $\, F_2$ and $\, F_3$ are remarkable second   
 and third order {\em globally nilpotent} linear
 differential operators~\cite{bo-bo-gu-ha-je-ma-ze-09,bo-bo-ha-ma-we-ze-09}.
The first order linear differential  operator $\tilde{L}_1$ quite  remarkably 
has a {\em polynomial solution}. 
The fifth order linear differential operator $\, L_5$ was shown to be 
equivalent  to the symmetric fourth power of (the second order operator) 
$L_E$ corresponding to the complete elliptic integral $\, E$.
The complete and detailed analysis of $\, L_{12}$, the order twelve
operator in (\ref{63}) is beyond our current
 computional ressources (see~\cite{bo-bo-gu-ha-je-ma-ze-09} for details).

It is important to note that these factorization results are exact and
have been obtained from series and ODEs obtained modulo a {\em single prime}.
For the reconstruction in exact arithmetics of the factors occurring in the
differential operator $\, L_{11}$, we had to obtain the series and ODEs for
more than one prime.
The length of the series necessary to obtain the underlying ODE
is initially unknown, except perhaps for some rough estimates.
Once the first non-minimal order ODEs have been obtained modulo a prime,
the {\em minimum length} of the series necessary to obtain non-minimal 
order ODEs for any other primes is {\em known exactly}.
This knowledge comes from a relation we reported  in~\cite{bo-gu-ha-je-ma-ni-ze-08} 
and  that we called the "ODE formula".
Beyond understanding the terms occurring in the "ODE formula" and the light they
shed on the ODEs underlying the problem, the formula has been of  most
importance in terms of gains in the computational effort.
For instance, we  initially generated, modulo a prime, 
10000 terms for $\, \tilde{\chi}^{(5)}$
and we found that we can obtain non-minimal ODEs using only some 7400 terms,
while non-minimal order ODEs for $\, \Phi^{(5)}$ can be obtained using some
6200 terms, representing a great reduction in the required computational effort.

\subsection{The ODE formula}

Let us denote by $Q$ the order of the ODE we are looking for and by $D$
the degree of the polynomials in front of the derivatives (we write the
ODE in the homogeneous derivative $x\, {d \over dx}$).
We must then have $(Q+1)(D+1)$ terms in the series in order to determine
the unknown polynomial coefficients.
If an ODE exists, it appears that the number of terms actually necessary
for the ODE to be obtained is given by
\begin{eqnarray}
\label{Q1D1}
N \,=\,\,  (Q+1)(D+1)\, -f,
\end{eqnarray}
where $f$ is a positive integer and indicates the number of ODE-solutions
to the linear system of equations for the polynomial coefficients.

From empirical observation, we have seen~\cite{bo-gu-ha-je-ma-ni-ze-08}
that $N$ is also given, linearly in terms of $Q$ and $D$, by  
\begin{eqnarray}
\label{empirical}
N \, = \, \, \,   
d \cdot Q \, \,  + q \cdot D \, \,  - C. 
\end{eqnarray}
While $Q$ and $D$ are the order and the degree, respectively,  of
 {\em any non-minimal order} ODE that  we choose to  look for, 
 the parameters $d$, $q$ and $C$ {\em depend} on the series we are
 working with.
In all the cases we have considered, we have found that $q$ is the order of the
{\em minimal order} ODE and $d$ is the number of singularities
(counted with multiplicity) excluding any apparent singularities
and the singular point $x=0$.
The parameter $C$ was shown in \cite{bo-bo-gu-ha-je-ma-ze-09}
to be in an exact relationship with the degree $D_{\rm app}$ of the apparent
polynomial of the minimal order ODE
\begin{eqnarray}
\label{Dapp}
D_{\rm app}  \,=\, \, (d-1)(q-1)\, -C\,  -1.
\end{eqnarray}

Note that there are many ODEs that annihilate a given series.
Among all these ODEs, there is a unique one of minimal order.
In our calculations we have seen that it is easier to produce  ODEs, which
are not of minimal order~\cite{ze-bo-ha-ma-05b},
in the sense that fewer terms are needed to obtain these ODEs 
compared to what is required to obtain the minimal order ODE. Even
more importantly for computational purposes, there is a non-minimal order
ODE that requires the minimum number of terms in order to be obtained.

Next we  demonstrate how we use the ODE formula to optimize our calculations,
i.e. generate just the necessary number of terms in the series.
From (\ref{Q1D1}-\ref{Dapp}), the parameter  $\, D$ is given as:
\begin{eqnarray}
\label{theD}
D \, =\, d\, -1 \, + {\frac{D_{\rm app}+f}{Q-q+1}}
\end{eqnarray}
and this must be a positive integer. The parameters $f$ and $Q$ are integers
with the constraints $f \, \ge\,  1$ and $Q\,  \ge\,  q$.
It is a simple calculation to run through the integers $\, f$ and $\, Q$ resulting in
a positive integer $D$. For each such triplet $(Q_0,\,  D_0,\,  f_0)$ the number
$N_0=\, (Q_0+1)(D_0+1)\,  -f_0$ is the number of terms in the series
required to obtain $f_0$ ODEs of order $Q_0$ and degree $D_0$.
Among all these $\, N_0$ there is a minimum. We call the corresponding ODE
the "optimal ODE". To obtain the ODE for other primes, it is thus only 
necessary to generate the minimum number of series terms.

For instance, for $\tilde{\chi}^{(5)}$, the ODE formula reads
\begin{eqnarray}
N \, = \, \, \,   
72 \cdot Q \, \,  + 33 \cdot D \, \,  -887 \, \,=  \,  \, (Q+1)(D+1)\, \,  -f.
\end{eqnarray}
The optimal ODE, i.e., the ODE that requires the minimum number of terms
in the series has the triplet $(Q_0, D_0, f_0)=\, (56, 129, 8)$ which
corresponds to the minimum number $N_0=\, 7402$.
Note that the minimal order ODE has the triplet $(33, 1456, 1)$ and requires
$49537$ series terms.

The minimum number of terms $\, N_0$ is implicitly given
by the ODE formula~(\ref{empirical}).
Plugging the parameter $D$ given in (\ref{theD}) in $N=\, (Q+1)(D+1)\, -f$, one
obtains
\begin{eqnarray}
\label{Nhyperbola}
N \, = \,\,  (Q+1)\,\,  d \, + D_{\rm app} \,
 + {\frac{ \left( D_{\rm app}+f \right)\, q }{Q-q+1}}.   
\end{eqnarray}
We can view $N$ as a {\em continuous} function of $Q$ and $f$ and we find
that it has two extremums when $\rmd\, N/\rmd Q =\, 0$. For
 the positive extremum one has
\begin{eqnarray}
&&Q_0 \, = \, q-1 + {\frac{1}{d}}\, \sqrt{(D_{app}+f)\,q\,d}, \\
&&D_0 \, = \, d-1 + {\frac{1}{q}}\, \sqrt{(D_{app}+f)\,q\,d}, \\
&&N_0 \, = \, q\,d + D_{app} + 2\, \sqrt{(D_{app}+f)\,q\,d}
\end{eqnarray}
For the example of $\tilde{\chi}^{(5)}$ considered above, one obtains
(with $f=1$)
\begin{eqnarray}
Q_0 \, \simeq  \,57.20, \quad D_0 \, \simeq \, 125.97, \quad N_0 \, \simeq \, 7388.09.
\end{eqnarray}

The gain in the number of terms is already very significant for $Q=q+1$ and can be
measured by the discrete derivative of the hyperbola $N(Q)$ given
in (\ref{Nhyperbola}). Since we should compute over the integers,
it is easier to compute the difference of $(D+1)\, (Q+1)$ evaluated
at the points $Q=q$ and $Q= \,q+1$. At the order $Q= \,q$, from (\ref{theD})
one obtains $D(Q=q)= \,d-1+D_{\rm app}+f_1$, where $f_1$ is a positive integer.
At the order $Q= \,q+1$, one has $D(Q=q+1)=\, d-1 \,+(D_{\rm app}+f_2)/2$, where $f_2$
is a positive integer with same parity as $D_{\rm app}$.
The gain in the number of terms is
\begin{eqnarray}
\label{DeltaN}
\Delta \, N(q, q+1) \, =\, \,-d \, + q\, f_1 \,
 + {1 \over 2} \, (D_{\rm app} - f_2)  \,q \,+ (f_1-f_2).
\end{eqnarray}
For $\chi^{(5)}$, and with the values $f_1=1$ and $f_2=2$
(since $D_{app}=1384$ is even), the ``saving'' in the number of terms is $22736$
 to be compared
with the $49537$ terms needed to obtain the minimal order ODE (i.e. $Q=q$).
As $Q$ increases, one approches the minimum of the hyperbola (\ref{Nhyperbola})
which is $N_0 \, = \, 7388.09$ (with $f=1$).
Over the integers the minimum is 7402 obtained with $f=8$.
This process can be repeated by computing $\Delta \, N(q, q+2)$
and in this case $D_{\rm app}+f_3$ should be multiple of 3.

As can be seen from the ``discrete" derivative (\ref{DeltaN}), the degree of
the apparent polynomial is crucial. For  ODEs with no apparent singularities
the minimal order ODE  {\em is}   the optimal ODE.
In this case, the hyperbola $N(Q)$ can still have a minimum that is not
in the integers.

Note that we may define a {\em minimal degree} ODE, i.e. the ODE that has
$D=d$ meaning that there is no singularities other than the ``true" singularities
of the minimal order ODE (no apparent and no spurious
singularities\footnote{If we denote by $L_q$ the {\em minimal order} differential
operator, the {\em non-minimal order} differential operator $L_{Q,D}$ (with
$Q >q$ and $D>d$) has $D-d$ singularities which are spurious with respect to
$L_q$. The spurious singularities are the ones of the operator $L_{Q-q}$ occurring in
the factorization $L_{Q,D}=\,L_{Q-q} \cdot L_q$.}).
The order of this {\em minimal degree} ODE is (see (\ref{theD}))
\begin{eqnarray}
Q \,\,  = \,\,\,  q\, + D_{app}\, +f\, -1,
\end{eqnarray}
giving for $ \, \tilde{\chi}^{(5)}$, the order $Q=1417$ and $103513$ as
the number of terms (the minimum $f$ being 1). 
Note that this {\em minimal degree} ODE is useless for our computational
purposes.

\vskip 0.2cm

In this paper all of these types of  modular calculations and approaches 
have been applied to $ \, \tilde{\chi}^{(6)}$.
Section~2 shows the computational details (timing, ...) for the generation
of the first series and the first ODEs, modulo a prime, from which we infer the optimal
length of the series to be generated for other primes. 
In Section~3, we report on the ODE annihilating $\, \tilde{\chi}^{(6)}$
and on the ODE annihilating the corresponding ``depleted" series.
The singularities and local exponents confirm the results obtained
from a diff-Pad\'e analysis and given in a previous paper~\cite{bo-gu-ha-je-ma-ni-ze-08}.
In Section~4, the program of factorization developed for $\, \tilde{\chi}^{(5)}$
is used to factorize as far as possible the differential operator
corresponding to the ODE of $\, \tilde{\chi}^{(6)}$.
We will see that our conjecture~\cite{bo-bo-gu-ha-je-ma-ze-09, ze-bo-ha-ma-05b}
on the factorization structure of the
$\, \tilde{\chi}^{(n)}$ holds for $n=6$.
Some right factors in the differential operator for $\, \tilde{\chi}^{(6)}$
are obtained in exact arithmetics.
Section 5 is the conclusion.

\section{The series of  $\tilde{\chi}^{(6)}$ modulo a prime}

As shown in~\cite{bo-gu-ha-je-ma-ni-ze-08} the calculation of a series
for $\tilde{\chi}^{(6)}$ is a problem with computational
complexity $O(N^4\ln N)$. Note that  $\tilde{\chi}^{(2n)}$
 is an even function in $w$ and we therefore
generally work with a series in the variable $x=\,w^2$, 
though the series for $\tilde{\chi}^{(6)}$ is still
calculated in the $w$ variable.
In Table~\ref{tab:ODEs} we have listed a summary of results for
the formula (11) for various series with new results for $\tilde{\chi}^{(6)}$ added.
In~\cite{bo-gu-ha-je-ma-ni-ze-08} we gave a rough estimate
of the number of terms required to obtain
 the ODE for $\tilde{\chi}^{(6)}$ and thought this beyond 
our computational resources. However, upon closer
 inspection of Table~\ref{tab:ODEs} one observes that
the minimum number of terms required to find the
 ODE in $x$ for $\tilde{\chi}^{(2n)}, \, n=1,\,2$ (or  $\Phi_H^{(2n)}$) is always
smaller than the number of terms required for
 $\tilde{\chi}^{(2n-1)}$ (or $\Phi_H^{(2n-1)}$). This also holds for the
combination $6\tilde{\chi}^{(n)}\,-(n-2)\,\tilde{\chi}^{(n-2)}$. It 
is reasonable to expect that this would be true for $\tilde{\chi}^{(6)}$
as well. In particular this  would mean that the number
 of terms required to find the ODE for $6\tilde{\chi}^{(6)}\,-4\tilde{\chi}^{(4)}$
should be smaller than the 6400 or so terms needed
 to find the ODE for $6\tilde{\chi}^{(5)}\,-3\tilde{\chi}^{(3)}$. 
There is of course no way of knowing whether 
or not this line of reasoning is correct. In particular we would
have liked to further reduce the number of terms 
to be calculated (one can for instance note that the number of
terms required to find the optimal ODE
 for $\tilde{\chi}^{(2n)}$ or $\Phi_H^{(2n)}$ is some 10-20\% less than
the number of terms required to find  the optimal ODE
 for $\tilde{\chi}^{(2n-1)}$ or $\Phi_H^{(2n-1)}$, respectively), 
but since finding the ODE for the first time is
 a hit-or-miss proposition we naturally wanted to ensure, to the greatest
extent possible, that we had enough terms 
 to find the ODE for $6\tilde{\chi}^{(6)}\,-4\tilde{\chi}^{(4)}$.
For this reason it was decided to generate
 a series to order 6500 in $x$ (13000 in $w$) for $\tilde{\chi}^{(6)}$
with the firm hope that this would suffice 
to find the optimal ODE for at least $6\tilde{\chi}^{(6)}\,-4\tilde{\chi}^{(4)}$ (in fact
it is also enough terms to find the optimal ODE for $\tilde{\chi}^{(6)}$  itself).

\begin{table}[htdp]
\caption{\label{tab:ODEs}
Summary of results for various series.   The last three columns are the data for the  optimal ODE. 
 The $\Phi_H^{(n)}$ series are the model integrals~\cite{bo-ha-ma-ze-07b}.
 }
\begin{center}
\begin{tabular}{|c|c|c|c|c|}\hline
    Series &  $N = d \cdot Q + q \cdot D - C $ &  $Q_0$ &   $D_0$ &      $(Q_0+1)(D_0+1)$ \\ \hline 
\hline
    $\tilde{\chi}^{(1)}$ &$1\, Q \, + \, 1\, D \, + 1$   &  1  &  1 &  4  \\
    $\tilde{\chi}^{(2)}$& $1\, Q \, + \, 2\, D \, + 1$   &  2  &  1 &  6  \\
    $\tilde{\chi}^{(3)}$& $12\, Q \, + \, 7\, D \, -37$  &  11 &  17&  216 \\
    $\tilde{\chi}^{(4)}$& $7\, Q \, + \, 10\, D \, -36$  &  15 &  9 &  160 \\
    $\tilde{\chi}^{(5)}$& $72\, Q \, + \, 33\, D \,-887$ & 56  & 129&  7410  \\
    $\tilde{\chi}^{(6)}$& $43\, Q \, + \, 52\, D \,-1121$& 84  &  73&  6290    \\
    $6\tilde{\chi}^{(3)}-\tilde{\chi}^{(1)}$&  $12\, Q \, + \, 6\, D \, -26$ & 10 & 17  &  198  \\
    $6\tilde{\chi}^{(4)}-2\tilde{\chi}^{(2)}$&  $6\, Q \, + \, 8\, D \, -17$ & 13 & 8   &  126  \\
    $6\tilde{\chi}^{(5)}-3\tilde{\chi}^{(3)}$& $68\, Q \, + \, 30\, D \,-732$& 52 & 120 &  6413 \\
    $6\tilde{\chi}^{(6)}-4\tilde{\chi}^{(4)}$& $40\, Q \, + \, 48\, D \,-945$& 80 &  66 & 5427  \\
    $\Phi_H^{(3)}$&  $10\, Q \, + \, 5\, D \, -21$     &  8  &  13  &  126     \\
    $\Phi_H^{(4)}$&  $5\, Q \, + \, 6\, D \, -12$      &  9  &  6   &  70      \\
    $\Phi_H^{(5)}$&  $45\, Q \, + \, 17\, D \, -277$   &  28 &  80  &  2349    \\
    $\Phi_H^{(6)}$&  $26\, Q \, + \, 27\, D \, -342$   &  48 &  39  &  1960    \\
    $\Phi_H^{(7)}$&  $145\, Q \, + \, 49\, D \, -1943$ &  92 &  257 &  23994   \\ \hline
\hline
 \end{tabular}
\end{center}
\end{table}

In~\cite{bo-gu-ha-je-ma-ni-ze-08} the calculation of $\tilde{\chi}^{(5)}$
to 10000 terms required some 17000 CPU hours on an
SGI Altrix cluster with 1.6GHz Itanium2 processors. Given that the algorithms
for  $\tilde{\chi}^{(5)}$ and  $\tilde{\chi}^{(6)}$ has the same
computational complexity this would indicate that the time required to
calculate the series for $\tilde{\chi}^{(6)}$ to 13000 terms in $w$ would
be at least 50000 CPU hours 
(the algorithm for $\tilde{\chi}^{(6)}$ has a slightly larger pre-factor
than that for $\tilde{\chi}^{(5)}$).
In fact it turned out that almost 65000 CPU hours was required and this
calculation was performed over a six months period.

The series to order 6500 was calculated modulo the prime 32749.
As in \cite{bo-bo-gu-ha-je-ma-ze-09} we want to factorise various
differential operators and reconstruct the right-most factors exactly
using the results from several primes.
We thus need to reduce as much as possible the length of the series
by identifying some right factors.

As we detail in the following section the optimal ODE for $\tilde{\chi}^{(6)}$ 
can be obtained with less than 6300 terms while the optimal ODE for the
combination $6\tilde{\chi}^{(6)}\,-4\tilde{\chi}^{(4)}$
requires `just' over 5400 terms.
Furthermore we find (using our series modulo a single prime)
that $\tilde{\chi}^{(2)}$ is a solution of this ODE and that one can
simplify further by considering the linear combination
$\Phi^{(6)}=\,\tilde{\chi}^{(6)}\,-\frac23 \tilde{\chi}^{(4)}\,+\frac{2}{45}\tilde{\chi}^{(2)}$
whose optimal ODE requires a little more than 5100 terms.

The ODE for $\tilde{\chi}^{(6)}$ has $\frac{{\rm d}^2}{{\rm d}x^2}$ as the
lower derivative, meaning that $c_1+c_2\,x$ is a solution ($c_1$ and $c_2$ are
constants).
Checking that $c_1+c_2\,x$ is still a solution of the ODE for $\Phi^{(6)}$
and producing the series $\frac{{\rm d}^2}{{\rm d}x^2} \Phi^{(6)}(x)$, we
arrive at a series whose minimal ODE requires a little
less than 5000 terms. 

We therefore calculated a further two series to order 5000 modulo the
primes 32719 and 32717. These calculations required an additional 45000
hours of CPU time. Using the factorisation procedure detailed in Section~4,
we found a  factor of order 3, $X_3$,  which right divides the differential
operator for $\frac{{\rm d}^2}{{\rm d}x^2} \Phi^{(6)}(x)$,  and we managed to 
reconstruct $X_3$ in exact arithmetic using 3 primes\footnote{$X_3$ is equivalent 
to the differential operator $L_3$ given in this paper.}.  Applying $X_3$,
that is form the series
 $X_3 \left(\frac{{\rm d}^2}{{\rm d}x^2} \Phi^{(6)}\right)$, results in a series 
whose optimal ODE requires less than 4800 terms. At
 about the same time as these developments took
place a new system was installed by National Computational Infrastructrure (NCI) whose 
National Facility provides the national peak computing facility for Australian researchers. 
This new system is an SGI XE cluster using quad-core 3.0GHz Intel Harpertown cpus. 
Our code runs  about 40\% faster (takes about $0.6$ times the time) on
this facility when compared to the Altix cluster
 and a calculation of a series to order 4800 takes
about 11000 CPU hours per prime. We  calculated series to this order for a further 6 primes,
namely, 32713, 32707, 32693, 32687, 32653 and 32647 (some of these calculations 
were performed  on the facilities of the Victorian Partnership for Advanced Computing using
a cluster with AMD Barcelona 2.3GHz quad core processors).

\section{Fuchsian differential equation for $\tilde{\chi}^{(6)}$}

From the $\tilde{\chi}^{(6)}$ series modulo a prime, we obtained various ODEs
which have the ODE formula
\begin{eqnarray}
\label{L52}
N \,\,=\,\,\, 43\, Q +52\, D\, -1121\, =\,\, (Q+1)(D+1)\,-f,
\end{eqnarray}
thus showing that the ODE for $\tilde{\chi}^{(6)}$ is of minimal order $52$.
We denote by $L_{52}$ the corresponding linear differential operator.

The polynomial in front of the highest derivative and carrying
the singularities of $L_{52}$ (i.e. the ODE of $\tilde{\chi}^{(6)}$)  reads
\begin{eqnarray}
&&(1\,-16\, x)^{30} \cdot (1\,-25\, x) \cdot (1\,-9\, x) \cdot
(1\,-\, x) \cdot (1\,-4\, x)^{5} \cdot (1\,-8\, x) \nonumber \\
&& \qquad \times (1\,-\, x +16 \, x^2) \cdot (1\,- 10\, x +29 \, x^2) \cdot P_{\rm app},
\end{eqnarray}
where $P_{\rm app}$ is a polynomial whose roots are apparent singularities.
Even though we have not computed the minimal order ODE, from (\ref{Dapp}), we
can infer that the degree of $P_{\rm app}$   is $D_{\rm app}=\,1020$.
All the singularities agree with the ones found in
\cite{bo-gu-ha-je-ma-ni-ze-08} from a diff-Pad\'e analysis and we have
confirmation that  $(1-8x)$  is the {\em only} singularity not predicted~\cite{bo-ha-ma-ze-07b}
by the simplified integrals $\Phi_H^{(6)}$.

Furthermore, using the exact (modulo a prime) ODE, we can confirm
the local exponents computed from a diff-Pad\'e analysis
in \cite{bo-gu-ha-je-ma-ni-ze-08} for all singularities except
those at $x=0, 1/16$ and $x=\infty$, which are correct but incomplete.
The complete set of local
 exponents\footnote[1]{The notation is $0^3$
 for $0, 0, 0$ and $9/2^2$ for $9/2, 9/2$, etc.}
at these latter points read:
\begin{eqnarray}
x=0, &\,\,&  \rho=\, -1, -1/2, 0^3, 1/2, 1^5, 3/2, 2^5, 3^4, 4^4, 5^4, 6^4,
7^3, 8^3, 9^3, \nonumber \\
&& \qquad 10^3, 11, \cdots, 17, \nonumber \\
x=1/16, &\,\,& \rho=\, -2, -7/4, -3/2, -5/4, -1^3, -1/2, 0^6, 1/2, 1^4,
2^4, 3^3, 4^3, \nonumber \\
&& \qquad 5^3, 6^2, 7^2, 8^2, 9^2, 10, \cdots, 21,  \nonumber \\
x=\infty, &\,\,& \rho = -1^2, -1/2, 0^3, 1/2^6, 1^2, 3/2^5, 2, 5/2^3,
3, 7/2^3, 4, 9/2^2, \nonumber \\
&& 11/2^2, 13/2^2, 15/2^2, 17/2^2, 19/2^2, 21/2^2,
23/2^2, 25/2, 27/2, \nonumber \\
&& 29/2, 31/2, 33/2, 35/2, 37/2, 19. \nonumber
\end{eqnarray}

Having obtained the ODE formula (\ref{L52}), one can see that the minimal
order ODE requires $56391$ terms
(plug $Q=\,q=\,52$, $d=\,43$, $D_{app}=\, 1020$ and 
$f=\,1$ into (\ref{Nhyperbola})).
And it is a simple calculation,
(see paragraph after (\ref{theD})) to obtain the number of terms necessary
for the optimal ODE. This corresponds to $Q=\,84$, $D=\,73$, $f=\,3$ and $N=\,6287$
terms.
If we had to produce the optimal ODE for $\chi^{(6)}$ for other primes
it is $6290$ series coefficients that should be generated.

As mentionned in the previous section, our conjecture that the
$\tilde{\chi}^{(n)}$ satisfy (with $\alpha_{n-2}=\,(n-2)/6$)
\begin{eqnarray}
\label{conject1}
\tilde{\chi}^{(n)} \,\,=\, \,\, 
\alpha_{n-2} \cdot  \tilde{\chi}^{(n-2)}\,\, \, 
+\beta_{n-4} \cdot  \tilde{\chi}^{(n-4)}\, 
+\, \cdots\,\, \, +\,  \Phi^{(n)},
\end{eqnarray}
is  also verified.
For the series
\begin{eqnarray}
\label{thePhi6}
\Phi^{(6)}\,\,=\,\,\,
\tilde{\chi}^{(6)}\,-\frac23 \tilde{\chi}^{(4)}\,+\frac{2}{45}\tilde{\chi}^{(2)},
\end{eqnarray}
we obtain  non-minimal order ODEs from which we infer the ODE formula
\begin{eqnarray}
39\, Q \,+ 46\, D \,- 861 \,=\,\, (Q+1)(D+1)\,-f, 
\end{eqnarray}
showing that the minimal order is 46 with an apparent
polynomial (see (\ref{Dapp})) of degree $D_{\rm app}=848$.
The minimal order ODE for $\Phi^{(6)}$ requires the generation of
$41736$ coefficients series,
while the optimal ODE requires $5120$ terms corresponding to $Q=\,79$, $D=\,63$.
It is interesting to see that the required number of terms
decreases sharply from $41736$ (for the minimal order ODE $Q=\,q=\,46$) to
$22272$ for the non-minimal order ODE $Q=\,q+1=\,47$.
The gain $\Delta N (46, 47)=\,19464$ terms
 is given by (\ref{DeltaN}) for $q=\,46$, $d=\,39$
$f_1=\,1$ and $f_2=\,2$ since $D_{\rm app}=\,848$ is even.
The gain $\Delta N (46, 48)\,=\,25958$.

Denoting by $L_{46}$ the differential operator corresponding to $\Phi^{(6)}$
and recalling~\cite{ze-bo-ha-ma-05b} the differential operator $L_{10}$
corresponding to $\tilde{\chi}^{(4)}$, one sees from (\ref{thePhi6}) that
the differential operator for $\tilde{\chi}^{(6)}$ has
the ``direct sum\footnote[5]{Recall
\cite{ze-bo-ha-ma-05b} that the differential operator for $\tilde{\chi}^{(2)}$
is a factor in the direct sum of $L_{10}$.} decomposition"
\begin{eqnarray}
L_{52} \, = \,\,\, L_{10} \oplus L_{46}.
\end{eqnarray}
The sum of the orders of the differential operators $L_{46}$ and $L_{10}$
is larger than 52, indicating that 
a common factor, namely an order four differential
operator, occurs at the right of both  $L_{46}$ and $L_{10}$.
The solutions of this order four ODE have been given in
eqs.~(31-33) and eq.~(43) ofé\cite{ze-bo-ha-ma-05b}.
The differential operator (that we denote by $\,L_4^{(4)}$) is given  
in eq.~(42) of~\cite{ze-bo-ha-ma-05b} as a product of four order one
differential operators. Since the expressions for these differential operators
were not  written in~\cite{ze-bo-ha-ma-05b},
 we give, for the sake of completeness,
in \ref{Ls4u4} the full factorization of the differential operator $L_4^{(4)}$.

Furthermore,   we note that in the ODE for $\tilde{\chi}^{(6)}$  the
derivatives of order zero and one are missing (the corresponding
differential operator has $D_x^2$ 
as the lowest derivative\footnote[2]{The notation
$D_x$ is ${\frac{d}{dx}}$.}).
The constant and the degree one polynomial $x$ are solutions of $L_{52}$.
The constant is a solution of the common factor $\,L_4^{(4)}$, but
the degree one polynomial $x$ is not a solution of  
$L_{10}$ and thus should occur in $L_{46}$.

We thus have  an order five differential operator that right divides $L_{46}$
\begin{eqnarray}
\label{common}
\tilde{L}_5 \, =\,\,   D_x^2 \oplus L_4^{(4)}\, 
 =\, \,\,  \left( D_x-{1 \over x} \right) \oplus L_4^{(4)}.
\end{eqnarray}

We now turn to the factorization modulo a prime of the differential
operator $L_{46}$ keeping in mind that $\tilde{L}_5$ is a right factor.

\section{Factorization modulo a prime of the differential operator $L_{46}$}

The local exponents at the singularities of the
ODE of $\Phi^{(6)}$ allows us to easily track the factors carrying the
various singular behaviours. What we mean is the following.
The local exponents for the ODE of $\Phi^{(5)}$, at for instance $w=\, 0$, are
all integers. Producing the series having the highest exponent,
we obtain either the full ODE or a right factor. If the series with the
highest exponent yields the full ODE then in order to obtain a right factor
we have to look at  the ODEs corresponding to combinations of series
involving both the highest and the
 next highest exponent as explained and done in~\cite{bo-bo-gu-ha-je-ma-ze-09}.

For $\Phi^{(6)}$ and at $x=0$, we have two types of local exponents, integer
and half-integer ones. We thus have  a ``partition" of the solutions to
the full ODE. In other words  we
have ``two highest exponents"\footnote{Note that for $\tilde{\chi}^{(5)}$,
other singularities than $w=\,0$ have half- and fourth-integers exponents.
There was no need in~\cite{bo-bo-gu-ha-je-ma-ze-09} to use the procedure
presented here.}
and it is therefore more likely that we can avoid using combination series.

The ODE for $\Phi^{(6)}$  corresponding to $L_{46}$ has at $x=\, 0$
the local exponents
\begin{eqnarray}
\rho \,&=&\,\, -1^2, -1/2, 0^3, 1/2, 1^5, 3/2, 2^5, 3^4, 4^4, 5^4, 6^3,
7^3, 8^3, 9^2, 10^2, 11, \nonumber \\
&& 12, 13     
\end{eqnarray}
we have then two ``highest exponents", $\rho=13$ and $\rho=3/2$.
This means that we can produce both the series $x^{\rho} \,(1+ \cdots )$
and see whether or not either of these gives rise to a right factor. If so
we may not need to resort to the combination method presented in
Section~4 of~\cite{bo-bo-gu-ha-je-ma-ze-09}.

At the singularity $x=1/16$, the local exponents are
\begin{eqnarray}
\rho \,&=&\,\,-2, -7/4, -3/2, -5/4, -1^3, -1/2, 0^6, 1/2, 1^4, 2^3, 3^3, 4^3,
5^2, 6^2, \nonumber \\
&& 7^2, 8, 9, \cdots,\, 19  \nonumber
\end{eqnarray}
and we have three ``highest exponents", $\rho=\,-5/4$, $\rho=\,1/2$ and $\rho=\,19$.

At the singularity $x=\infty$, there are two ``highest exponents", $\rho=\,4$
and $\rho=\,33/2$ since the local exponents are
\begin{eqnarray}
\rho \,&=&\,-1^2, -1/2^2, 0^3, 1/2^6, 1^2, 3/2^5, 2, 5/2^2, 3, 7/2^2,
4, 9/2^2, 11/2^2, \nonumber \\
&& 13/2^2, 15/2^2, 17/2^2, 19/2^2, 21/2, \cdots,\, 33/2
\nonumber
\end{eqnarray}

Before we proceed, we introduce the notation $L_{46} =\, O_{n_2} \cdot O_{n_3}$,
with $46=\,n_2+n_3$, which we use to indicate that the operator $L_{46}$ factorizes
into two operators of orders $n_2$ and $n_3$, respectively. Only when a
differential operator is definitive do we give it a label other than the $O$.

Let us begin by the conjecture \cite{bo-bo-gu-ha-je-ma-ze-09} that $L_{46}$ 
has a left-most operator of order six which is the symmetric fifth power of $L_E$.
Solutions to the symmetric power of $L_E$ are polynomials of homogeneous 
degrees in the elliptic integrals with the
 coefficients of the combination being rationals.
The solutions carrying the half-integer exponents
 should therefore be those of an operator
occurring necessarily at the right of $L_{46}$. So
 from the two ``highest exponents" $\rho=\,13$ 
and $\rho=\,3/2$ at $x=\,0$,  we need only obtain the ODE of the unique
series $x^{3/2}\, (1+ \cdots)$.
Indeed, acting by $L_{46}$ on the series $x^{3/2} \cdot (1+\cdots )$ produces
a series annihilated by an order 40 ODE, leading to the factorization
\begin{eqnarray}
\label{factoL6}
L_{46} \,\, =\,\,\, L_6 \cdot O_{40}.
\end{eqnarray}

When we shift $L_{46}$ to $x=\,1/16$ and act on $t^{-5/4} \cdot (1+ \cdots)$,
with $t=\,x-1/16$, 
we obtain an order five ODE leading to:
\begin{eqnarray}
L_{46} \,\, =\,\,\, O_{41} \cdot O_5.
\end{eqnarray}
Shifting $L_{46}$ to $x=1/16$ and acting on $t^{1/2} \cdot (1+ \cdots)$
produces:
\begin{eqnarray}
L_{46} \,\, =\,\, \,O_{36} \cdot O_{10}.
\end{eqnarray}
Shifting $L_{46}$ to $x=\,\infty$ and acting on $t^{4} \cdot (1+ \cdots)$,
with $t=\,1/x$, gives:
\begin{eqnarray}
L_{46} \,\, =\,\,\, O_{33} \cdot O_{13}.
\end{eqnarray}
Some factors are common to these   three factorizations. Shifting the
ODE back to $x=\,0$ and carrying 
out our factorization procedure~\cite{bo-bo-gu-ha-je-ma-ze-09},
one obtains (some final labelling is given) 
\begin{eqnarray}
L_{46} \,\, =\,\,\, O_{41} \cdot O_{5}
\, =\,\, O_{41} \cdot \tilde{L}_3 \cdot L_2,
\end{eqnarray}
\begin{eqnarray}
L_{46} \,\, =\,\, \,O_{36} \cdot O_{10}
 \,=\,\, O_{36} \cdot O_{1} \cdot L_4 \cdot \tilde{L}_3 \cdot L_2,
\end{eqnarray}
\begin{eqnarray}
\label{factoL4L3L2}
L_{46} \,\, =\,\,\, O_{33} \cdot O_{13}
\, =\,\, O_{33} \cdot O_{1} \cdot O_3 \cdot L_4 \cdot \tilde{L}_3 \cdot L_2.
\end{eqnarray}
The order one differential operator $O_1$ in the last factorization is equivalent
to an order one differential operator 
occurring in the $\tilde{L}_5$ of (\ref{common}).
The product  $O_3 \cdot L_4 \cdot \tilde{L}_3 \cdot L_2$ can be expressed as
a direct sum
\begin{eqnarray}
\label{factoL3oplus}
O_3 \cdot L_4 \cdot \tilde{L}_3 \cdot L_2\, =\,\,\,
L_3 \oplus \left( L_4 \cdot \tilde{L}_3 \cdot L_2  \right).
\end{eqnarray}

Collecting the results given in the factorizations (\ref{factoL6}) and
(\ref{factoL4L3L2}) with (\ref{factoL3oplus}), and keeping in mind the
right factor (\ref{common}), one obtains
\begin{eqnarray}
\label{lastfacto}
L_{46} \,\, =\,\,\, L_6  \cdot L_{23}  \cdot L_{17}
\end{eqnarray}
with
\begin{eqnarray}
L_{17} \, &=&\,\,\,
 \tilde{L}_5 \oplus  L_{3} \oplus \left(  L_4 \cdot \tilde{L}_3 \cdot L_2  \right), \\
\tilde{L}_5 \,&=&\,\, \,\left( D_x-{1 \over x} \right) \oplus L_4^{(4)}.
\end{eqnarray}

Having obtained all these differential operators,
 a final check is performed by acting on
$\Phi^{(6)}$ by the corresponding ODEs in the order given in (\ref{lastfacto})
and doing this we do indeed get zero.

\subsection{The differential operator $L_6$}

The sixth  order linear differential operator $L_6$  is the one that we
conjectured~\cite{bo-bo-gu-ha-je-ma-ze-09}  should annihilate a homogeneous 
polynomial of the complete elliptic integrals $\, E$ and $\, K$
of (homogeneous) degree five. It should then be irreducible.
The local exponents at the origin of the linear 
ODE corresponding to $L_6$ are
\begin{eqnarray}
x =\,  0, \quad \quad \quad  \quad 
\rho\,=\, \,-12,\,\,-11,\, \,-8, \,\,-5, \,\,-4, \,\, 0  
\end{eqnarray}
Plugging a generic series $\sum c_n\, x^n$  into the linear ODE fixes all the
coefficients with the exception of the coefficient $c_{0}$. 
The ``survival" of   a single coefficient is a particular feature 
of an irreducible factor with one non-logarithmic solution.
The differential operator $L_6$ being a symmetric power of $L_E$ means that
its solution is a polynomial in $E$ and $K$ defined as
\begin{eqnarray}
K  \, =\, _2 F_1 \left( [1/2, 1/2], [1], 16x \right), \quad
E  \, =\, _2 F_1 \left( [1/2, -1/2], [1], 16x \right).
\end{eqnarray}
The ODE corresponding to $L_6$ should only have singularities at $x= \, 0, 1/16$
and $x=\, \infty$, and this is indeed the case. The
 local exponents at $x=\, 1/16$ are
\begin{eqnarray}
x =\,  1/16, \quad \quad \quad  \quad 
\rho\,=\, \,-48^2,\,\,-47,\, \,-44, \,\,-40, \,\,0.  
\end{eqnarray}

The local exponents at $x=0$ and $x=1/16$ suggest
the following ansatz to be plugged into the linear ODE (of $L_6$):
\begin{eqnarray}    
\label{eqKE}
{\frac{1}{x^{12} \cdot  (1-16x)^{48}}} \cdot \, 
 \sum_{i=0}^5 \, P_{5-i, i}(x) \cdot  K^{5-i} \, E^i.
\end{eqnarray}
The polynomials  $P_{5-i, i}(x)$ can be determined numerically and 
the solution (analytical at $x=0$) of the ODE corresponding to $L_6$ is
\begin{eqnarray}
{\frac{1}{x^{12} \cdot (1-16x)^{48}}} \cdot \,&&
\Bigl( (1-16x)^4 \, P_{5,0} \cdot  K^{5} \, 
+(1-16x)^3 \, P_{4,1}\cdot  K^4\,E \nonumber \\
&& + (1-16x)^2  \, P_{3,2}\cdot  K^3 \, E^2
 + (1-16x) \, P_{2,3} \cdot K^2 \,E^3 \nonumber \\
&&  + P_{1,4} \cdot K \,E^4  \, + P_{0,5} \cdot E^5 \Bigr).
 \nonumber
\end{eqnarray}
The polynomials $P_{5-i, i}(x)$ with coefficients known modulo a
prime, are of degree respectively, 111, 112, 113, 113, 113 and 113.
As conjectured
the linear differential operator $L_6$ {\em is thus equivalent to 
the symmetric fifth power of} $L_E$.

\subsection{The differential operator $L_{17}$}

The differential operator $L_{17}$ has in its decomposition the differential
operator $\tilde{L}_5$ which is known exactly. The solutions of $\tilde{L}_5$
are the degree one polynomial $x$ and the four solutions of $L_4^{(4)}$ given
in \cite{ze-bo-ha-ma-05b}.
As for the other factors of $L_{17}$, i.e. $L_2$, $L_3$, $\tilde{L}_3$ and $L_4$, 
we have been able to express all of them in exact
arithmetics.

To express a differential operator  in
 exact arithmetics the straightforward approach
is to rationally reconstruct the differential
 operator using several modulo prime calculations.
However, an alternative would be to reconstruct the solutions to the 
differential operator if they are known. This is
 what we have done for $L_2$ and $L_3$.

The singularities of the ODEs corresponding to $L_2$ and $L_3$ are
{\em only} $x=0$, $x=1/16$ and $x=\infty$.
It is therefore reasonable to assume that the solutions can be expressed
as polynomials in $K(x)$ and $E(x)$.

For the ODE corresponding to $L_2$, the solution (analytical at $x=\, 0$)
written in terms of $\tilde{\chi}^{(2)}$, is:
\begin{eqnarray}
{\rm sol}(L_2)  \, =\,\,
 \left( x {\rmd \over \rmd x} -2 \right) \, \tilde{\chi}^{(2)}.
\end{eqnarray}
Written in this way,  it is easy to recognize the coefficients
in exact arithmetics with only two primes.
The differential operator $L_2$ is thus:
\begin{eqnarray}
L_2 \,\, =  \, \,\,
 D_x^{2}\, -2\,{\frac { \left( 1+8\,x \right) }{x \cdot (1-16\,x) }}\, D_x 
\, \,  +{\frac {4}{x \cdot  \left( 1-16\,x \right) }}.
\end{eqnarray}

For the third order differential operator $L_3$, we assumed
that it is equivalent to a symmetric square of $L_E$.
Indeed, the solution (analytical at $x=\, 0$)
written also in terms of $\left( \tilde{\chi}^{(2)} \right)^2$,
appears as:
\begin{eqnarray}
\fl \qquad {\rm sol}(L_3)  \, =\,\, {1 \over x} \cdot
\Bigl( x\,(1-16x)^2\,(16x-3)\cdot {\rmd^2 \over \rmd x^2} \,
+(1-16x)\,(64x^2-44x+9)\cdot {\rmd \over \rmd x} \nonumber \\
\fl \qquad \qquad \qquad \,
 -8\, (1-8x)\,(16x+9) \Bigr) \, \left( \tilde{\chi}^{(2)} \right)^2.
\end{eqnarray}
Here also, two primes are more than sufficient to recognize the
coefficients. The differential operator $L_3$, in exact arithmetics, reads
\begin{eqnarray}
L_3 \,\,=\,\,\, D_x^3\,  \, + {\frac{p_2}{p_3}}\,D_x^2\,\,  +{\frac{p_1}{p_3}}\,D_x
\,\,  +{\frac{p_0}{p_3}},
\end{eqnarray}
with:
\begin{eqnarray}
\fl \qquad p_3 \,=\,\, 
{x}^{2} \cdot (1-16\,x)^{2} 
\left(  -81+1986\,x-17056\,{x}^{2}+34304\,{x}^{3}+8192\,{x}^{4} \right), 
\nonumber \\
\fl \qquad p_2  \,=\,\, 
2\,{x}^{2} \cdot (1-16\,x)  \, ( 2247-46496\,x+357888\,{x}^{2}
-565248\,{x}^{3}-65536\,{x}^{4}), \nonumber \\
\fl \qquad p_1  \,=\,
6\,\left( 27-942\,x+11152\,{x}^{2}-101632\,{x}^{3}+372736\,{x}^{4}
-65536\,{x}^{5} \right), \nonumber \\
\fl \qquad p_0  \,=\,\, 
12\, (9-308\,x-6208\,{x}^{2}-101376\,{x}^{3}-49152\,{x}^{4}). \nonumber
\end{eqnarray}

We have not been able to find the solution
 of the ODE coresponding to $\tilde{L}_3$.
The rational reconstruction has been done on the differential
operator itself (see \ref{apptildeL3}).
Rationally reconstructed, the differential operator $\tilde{L}_3$ reads
\begin{eqnarray}
\label{tildeL3}
\tilde{L}_3\,  \,=\,\,\,   D_x^3\, \,  + {\frac{q_2}{q_3}}\,D_x^2
\,\,  +{\frac{q_1}{q_3}}\,D_x\,\,  +{\frac{q_0}{q_3}},
\end{eqnarray}
with:
\begin{eqnarray}
q_3 &=&\, 
{x}^{2} \cdot (1-4\,x) \left( 1-16\,x \right)^{3}\, Q_3, \nonumber \\
Q_3 &=&\,   -8+252\,x-1678\,{x}^{2}+3607\,{x}^{3} +4352\,{x}^{4}, 
  \nonumber \\
q_2  &=&
2\,x \cdot (1-16\,x)^{2}
\Bigl( -12 +1172\,x -30499\,{x}^{2} +252146\,{x}^{3} \nonumber \\
&& -872579\,{x}^{4}+770128\,{x}^{5} +1183744\,{x}^{6} \Bigr), 
  \nonumber \\
q_1  &=&\, 
4\,\,  (1-16\,x)  \Bigl(6 +185\,x-28373\,{x}^{2}+689440\,{x}^{3}
-5128290\,{x}^{4}  \nonumber \\
&& +16119599\,{x}^{5} -13139200\,{x}^{6}-17825792\,{x}^{7} \Bigr), 
  \nonumber \\
q_0  &=&\, 
4\,\, \Bigl( -294+9469\,x+84480\,{x}^{2}-4652220\,{x}^{3}
+33948640\,{x}^{4}  \nonumber \\
&& -97687536\,{x}^{5}+89128960\,{x}^{7}+74981376\,{x}^{6} \Bigr).
 \nonumber
\end{eqnarray}

All the calculations on the previous differential operators have been done
with the two primes 32749 and 32719. For the differential operator $L_4$
we need more primes.
The differential operator  $L_4$ has the form
\begin{eqnarray}
\label{theL4}
\fl \qquad \qquad  L_4 \, \,=\, \, \, 
{x}^{3} \cdot (1-16\,x)^{4} \left( 1-4\,x \right) 
 \, (1-8\,x) \,\,  Q_3^{4} \, P_4^{(26)} \cdot D_x^4 \nonumber \\
\fl \qquad \qquad \qquad 
+ {x}^{2} \cdot (1-16\,x)^{3}\,  \, Q_3^3 \, P_3^{(33)} \cdot D_x^3
\, + x \left( 1-16\,x \right)^{2} \,\,  Q_3^2 \,  P_2^{(38)} \cdot D_x^2 
\nonumber \\
\fl \qquad \qquad \qquad 
 + \, (1-16\,x) \, Q_3\, P_1^{(43)} \cdot D_x
\,\,  + P_0^{(47)},
\end{eqnarray}
where $Q_3$ is the apparent polynomial of $\tilde{L}_3$ in (\ref{tildeL3})
and $P_j^{(n)}$ are polynomials in $x$ of degree $n$.
To perform the rational reconstruction of the polynomials $P_j^{(n)}$,
we had to generate the series for $\Phi^{(6)}$ for another seven primes, then
obtain the optimal ODEs and factorize  the differential
operators $L_{46}$ for each prime. After the rational reconstruction was 
completed successfully the resulting differential
operator $L_4$ was checked against the local exponents and the
conditions on the apparent singularities.
The polynomials $P_j^{(n)}$ are given in exact arithmetics in
\ref{appL4}.

Note that we have also checked that these rationally reconstructed differential
operators are globally nilpotent as they should be.

\subsection{The differential operator $L_{23}$ }

The differential operator $L_{23}$ has the ODE formula
\begin{eqnarray}
21 Q\,  + 23 D\,  +1360\,  =\, \,  (Q+1)(D+1)\, -f,
\end{eqnarray}
and at $x=0$, the local exponents read:
\begin{eqnarray}
\rho \,&=&\,-25, -24, -23^2, -22^2, -21^2, -20^2, -19, -18, -17^2, -16, 1,
2, 3, \nonumber \\
&& 4, 5, 6, -47/2, -45/2 \nonumber
\end{eqnarray}
We can use the same method as before in order to factorize $L_{23}$.
By producing the series with the 
highest local exponents $\rho=\, 6$ and $\rho=\, -45/2$,
we obtained the full ODE for each series, i.e. an
ODE formula compatible with the minimal order 23.

The singularities of the linear ODE corresponding to $L_{23}$
are (besides $x=\, 0$): 
\begin{eqnarray}
\fl \quad (1-16\,x) \, (1-4\,x)  \, (1-x) \, (1-9\,x)
 \, (1-25\,x)  \, (1-10\,x+29\,{x}^{2}) 
  \, (1-x+16\,{x}^{2})   \nonumber
\end{eqnarray}
We may then shift the ODE corresponding to $L_{23}$ to a singular point other
than $x=\,0$, produce the series of the highest exponent and see whether
this gives an ODE of order less than 23.
At $x=\,1/16$, the series of the highest exponent $\rho=\,11$  produced the
full ODE.
Likewise, at other points and exponents
 such  as $(x=\,1/4,  \, \rho=\,-41/2)$,
 $(x=\, 1/9, \,  \rho=\,-47/2)$,
$(x=\,1/25,  \, \rho=\,-63/2)$, $(x=\,1, \, \rho=\,-47/2)$ 
and $(x=\,\infty, \,  \rho=\,-38,\, -47/2)$,
the   series   give rise to the full ODE.
 
Next we show how the {\em local structure} of solutions appear around $x=\,0$.
We introduce the notation $[x^p]$ to mean that the series
 begins as $x^p \cdot (const.\, + \cdots)$.
The results of our computations are the following.
Two sets of five solutions can be written as (with $k=1,2$)
\begin{eqnarray}
\label{series1}
\fl \qquad \quad 
 [x^k] \, \ln(x)^4\, + [x^{-21}] \, \ln(x)^3 \,+ [x^{-22}] \, \ln(x)^2 \,+
[x^{-23}] \, \ln(x) \, + [x^{-25}],  \,  \nonumber \\
\fl \qquad \quad   [x^k] \, \ln(x)^3\, + [x^{-21}] \, \ln(x)^2 \,+
[x^{-22}] \, \ln(x) \, + [x^{-24}],  \,  \nonumber \\
\fl \qquad \quad   [x^k] \, \ln(x)^2\, + [x^{-21}] \, \ln(x)\, + [x^{-24}],  
 \nonumber \\
\fl \qquad \quad   [x^k] \, \ln(x) \, + [x^{-21}] 
\qquad  \hbox{and}   \qquad  [x^k]. 
\end{eqnarray}
Three sets of three solutions can be written as (with $k=\,3,\, 4,\, 5$)
\begin{eqnarray}
\label{series2}
\fl \qquad \quad  
 [x^k] \, \ln(x)^2\, +[x^{-21}] \, \ln(x) \,+ [x^{-24}], 
 \nonumber \\
\fl \qquad \quad   [x^k] \, \ln(x) \, + [x^{-21}]
\qquad  \hbox{and} \qquad  [x^k].  
\end{eqnarray}
Two solutions can be written as
\begin{eqnarray}
\label{series3}
\fl \qquad \quad    [x^6] \, \ln(x) \, + \,   [x] \qquad  \hbox{and} \qquad  [x^6]  
\end{eqnarray}
Finally there are two non-logarithmic solutions
 behaving as $x^{-47/2}\cdot (1+\cdots)$ and
$x^{-45/2} \cdot (1 \, +\cdots)$.

Besides the series $x^\rho \cdot (1\, + \cdots)$ with  ($\rho=\, 6$ and $\rho=\, -45/2$)
that have given the full ODE, we may even try the non ambiguous solutions
such $[x^2]$ in front of $\ln(x)^4$ and $[x^5]$ in front of $\ln(x)^2$.
But these series produce the full ODE.

As is the case with the twelfth order differential
 operator $L_{12}$ occurring in $\tilde{\chi}^{(5)}$,
we have no final conclusion as to whether or not $L_{23}$ 
is reducible, and without performing
the factorization based on the combination method presented
 in Section~4 of~\cite{bo-bo-gu-ha-je-ma-ze-09}
 we do not expect to be able to reach any such conclusion.
The representative optimal ODE of $L_{23}$ used in the calculations is of
order 67, making the computational time obstruction more severe than what we
faced with the twelfth order differential operator
 occurring~\cite{bo-bo-gu-ha-je-ma-ze-09}
in $\tilde{\chi}^{(5)}$.

\subsection{Summary}

Let us now summarize our results.
The linear differential operator $L_{46}$, corresponding to 
$\Phi^{(6)}\, =\, \, \tilde{\chi}^{(6)}\,  - {\frac{2}{3}} \tilde{\chi}^{(4)}\, 
+{\frac{2}{45}}  \tilde{\chi}^{(2)}$ can be written as
\begin{eqnarray}
L_{46} \,=\,\, \,\, L_6 \cdot L_{23} \cdot L_{17},
\end{eqnarray}
with
\begin{eqnarray}
L_{17} \,\,  =\,\,\,   L_4^{(4)} \oplus \left( D_x-{1 \over x} \right) \oplus  L_{3}
\oplus  \left( L_4 \cdot \tilde{L}_3 \cdot L_2  \right)
\end{eqnarray}
The order seventeen linear differential operator $L_{17}$ contains only  the
   singularities of   the linear ODE corresponding to $L_{10}$
 (the operator for $\tilde{\chi}^{(4)}$) plus
the ``new''\footnote[3]{It is ``new" with respect to what we  obtained
from the $\Phi_H^{(6)}$ integrals~\cite{bo-ha-ma-ze-07b} and our Landau 
singularity analysis~\cite{bo-gu-ha-je-ma-ni-ze-08}.} 
 singularity $x= \, 1/8$.
 The singularity $x= \, 1/8$ occurs only in the fourth order linear
 differential operator $L_4$.
The third order differential operator $\tilde{L}_3$ 
is {\em responsible for the} $\rho=\, -5/4$,
$\rho= \, -7/4$ {\em singular behavior} around the
 (anti-)ferromagnetic point $x=\, 1/16$.

Comparing the results of  $\tilde{\chi}^{(6)}$ with those of
$\tilde{\chi}^{(3)}$, $\tilde{\chi}^{(4)}$ and $\tilde{\chi}^{(5)}$
we note that our conjecture   still holds: for a given $\tilde{\chi}^{(n)}$
there is an order $n$ differential operator equivalent to the $(n-1)$-th
symmetric power of
$L_E$ at left of the depleted differential operators, corresponding
to the linear combinations 
$\tilde{\chi}^{(3)}\,  - {\frac{1}{6}} \tilde{\chi}^{(1)}$,
$\tilde{\chi}^{(4)}\,  - {\frac{2}{6}} \tilde{\chi}^{(2)}$,
$\tilde{\chi}^{(5)}\,  - {\frac{3}{6}} \tilde{\chi}^{(3)}\,
 + {\frac{1}{120}} \tilde{\chi}^{(1)}$
and now
$\tilde{\chi}^{(6)}\,  - {\frac{4}{6}} \tilde{\chi}^{(4)}\, 
+{\frac{2}{45}}  \tilde{\chi}^{(2)}$.

For a given $\tilde{\chi}^{(n)}$ and once the ``contributions" of lower
terms ($\tilde{\chi}^{(n-2k)}$, $k=\, 1, \, 2,\,  \cdots$) 
have been substracted, the ODE of the
``depleted" series still contains some factors occurring in the ODE
of the lower terms ($\tilde{\chi}^{(n-2k)}$).
For $\tilde{\chi}^{(5)}$, we have that the differential operator $Z_2\cdot N_1$,
which occurs in the ODE of $\tilde{\chi}^{(3)}$, continues to be a right factor
in the ODE of
 $\tilde{\chi}^{(5)}\,  - {\frac{3}{6}} \tilde{\chi}^{(3)}\,
 + {\frac{1}{120}} \tilde{\chi}^{(1)}$.
For $\tilde{\chi}^{(6)}$, we have that the differential operator $L_4^{(4)}$,
which occurs in the ODE of $\tilde{\chi}^{(4)}$, continues to be a right factor
in the ODE of
$\tilde{\chi}^{(6)} - {\frac{4}{6}} \tilde{\chi}^{(4)}
+{\frac{2}{45}}  \tilde{\chi}^{(2)}$.

As was the case for $\tilde{\chi}^{(5)}$ with the differential operators
of order two and three ($F_2$ and $F_3$), we similarly have for
$\tilde{\chi}^{(6)}$, the emergence of two differential
operators of order three and four ($\tilde{L}_3$ and $L_4$), which are
globally nilpotent and for which we have no solutions.
We may imagine that all these ODEs have solutions in terms
(of symmetric power) of hypergeometric functions (with pull-back) as 
we succeeded to show~\cite{bo-bo-ha-ma-we-ze-09} for $Z_2$.
Providing these solutions in terms of modular forms is clearly a
challenge.

Similarly to the twelfth order differential operator $L_{12}$ occurring in
$\, \tilde{\chi}^{(5)}$, we faced with the differential operator $L_{23}$
the same obstruction to its potential factorization,
 namely prohibitive computational times.

\section{Conclusion}

We have calculated, modulo a prime, a long series for the six-particle
contribution $\tilde{\chi}^{(6)}$ to the magnetic susceptibility of the
square lattice Ising model.
This series has been used to obtain the Fuschian differential equation
that annihilates $\tilde{\chi}^{(6)}$. 

The method of factorization~\cite{bo-bo-gu-ha-je-ma-ze-09}
previously used for $\tilde{\chi}^{(5)}$ is
applied to the differential operator $L_{52}$ of $\tilde{\chi}^{(6)}$.
With the ODE known modulo a single prime, we have been able to go, as far as
the computational ressources allow, in the factorization of the
corresponding differential operator.

We have found several remarkable results.
The factorization structure  of $L_{52}$ generalizes what we have found for 
the linear differential operators of $\, \tilde{\chi}^{(3)}$, $\, \tilde{\chi}^{(4)}$
and $\, \tilde{\chi}^{(5)}$.
In particular, we found in $\, \tilde{\chi}^{(6)}$ the occurrence of the
term $\, \tilde{\chi}^{(4)}$ but also the lower term $\, \tilde{\chi}^{(2)}$,
leading to the differential operator $L_{46}$ corresponding to the
``depleted" series
 $\Phi^{(6)}\,=\,\,\tilde{\chi}^{(6)}\, - {\frac{2}{3}}\, \tilde{\chi}^{(4)}\,
+{\frac{2}{45}}  \tilde{\chi}^{(2)}$.
The left-most factor $L_6$ of $L_{46}$ is a sixth order operator
equivalent to the symmetric fifth power 
of the second  order operator $L_E$ corresponding to
complete elliptic integrals of the first (or second) kind.
We expect that this happens for  all  $\, \tilde{\chi}^{(n)}$, i.e. we conjecture the
occurrence in $\, \tilde{\chi}^{(n)}$ of terms proportional to $\, \tilde{\chi}^{(n-2k)}$
meaning a direct sum structure, and the occurrence of a $n-$th order
differential operator that left divides
the differential operator corresponding to the ``depleted" series
(\ref{conject1}) of $\, \tilde{\chi}^{(n)}$.

Some right factors of small order appear in the factorization of $L_{46}$.
We have used the previously reported ``ODE formula" to optimize our calculations.
We have generated other series of the {\em minimum number of terms}, modulo
eight other primes, and have obtained the corresponding ODEs and the corresponding
factorizations.
These nine factorizations have been used to perform a rational
reconstruction and obtain in exact arithmetics the
right factors occurring in $L_{46}$.

Our analysis is lacking the factorization of $L_{23}$ for which, and
similarly to $L_{12}$ occurring in $\, \tilde{\chi}^{(5)}$, we have
no conclusion on whether they are reducible.
Even if these differential operators are known in exact arithmetics,
their factorization remains a challenge for the methods implemented in
various packages of symbolic calculation.

The massive calculations performed on $\, \tilde{\chi}^{(5)}$ and
$\, \tilde{\chi}^{(6)}$ are at the limit of our computational ressources
and the next step, namely  $\, \tilde{\chi}^{(7)}$ and/or  $\, \tilde{\chi}^{(8)}$
seems to be really out of reach. 
A motivation for obtaining these very high order Fuchsian operators is to
understand  hidden mathematical structures from the factors of these
operators. In this respect, the main results we have obtained
on $\tilde{\chi}^{(6)}$ are the order three and four operators
($\tilde{L}_3$ and $L_4$) that we succeeded to get in exact arithmetics
and which are waiting for an {\em elliptic curve} mathematical interpretation.
Providing a mathematical interpretation for all these differential operators
in terms of modular forms is clearly our next challenge.

The   series and  differential operators studied
 in this paper can be found at~\cite{http}.

\ack
We are grateful to A. Bostan for checking the global nilpotence of the
rationally reconstructed differential operators $\tilde{L}_3$ and $L_4$.
IJ is supported by the Australian Research Council under grant DP0770705.
The calculations  would not have been possible 
without a generous grant from the National 
Computational Infrastructure (NCI) whose National Facility 
provides the national peak computing facility for Australian researchers.
We also made use of the facilities of the Victorian Partnership for Advanced Computing (VPAC). 
This work has been performed without any support of the ANR, the ERC, the MAE.

\appendix

\section{The order four differential operator $L_4^{(4)}$}
\label{Ls4u4}
The order four differential operator $L_4^{(4)}$ is a right factor 
in $L_{10}$ the differential operator for $\tilde{\chi}^{(4)}$.
It is a product of an order one differential operator and an order
three differential operator that can be written as a direct sum:
\begin{eqnarray}
L_4^{(4)} \,\, =\,\,\,
\, L_{1,3}^{(4)} \cdot \left( L_{1,2}^{(4)} \oplus L_{1,1}^{(4)} \oplus D_x \right)
\end{eqnarray}
In terms of the variable $x=w^2$ they are:
\begin{eqnarray}
L_{1,1}^{(4)} \,\,=\,\,\,
 D_x\,\,+ {\frac {768 \, {x}^{2}}{ (1-16\,x)  \, (1-24\,x+96\,{x}^{2}) }},
\end{eqnarray}
\begin{eqnarray}
L_{1,2}^{(4)} \,\,=\,\,\,
 D_x\,\,+{\frac {1+384\,{x}^{2}+2048\,{x}^{3}}{2\,x \cdot
 (1-16\,x)  \, (1-48\,x+128\,{x}^{2}) }},
\end{eqnarray}
and
\begin{eqnarray}
L_{1,3}^{(4)}\, \,=\,\, \, D_x\,\, +2\,{\frac {p_0}{p_1}}.
\end{eqnarray}
with:
\begin{eqnarray}
\fl \quad p_1 \,=\,\, x \cdot (1-16\,x) \, (1- 4\,x)
\, (80\,x+7)  \, (-7+96\,x-1152\,{x}^{2}+10240\,x^3),  
   \nonumber \\
\fl \quad p_0 \,=\,\,
 65536000\,{x}^{6}-36536320\,{x}^{5}
+481280\,{x}^{4}+254592\,{x}^{3}-24800\,{x}^{2}+2149\,x-49.
 \nonumber 
\end{eqnarray}

\section{Reconstruction in exact arithmetics of the differential operator $\tilde{L}_3$}
\label{apptildeL3}
The ODE corresponding to $\tilde{L}_3$ appears as (where the singularities
are easily recognized):
\begin{eqnarray}
\tilde{L}_3\,\, &=&\,\,
x^{3} \cdot (x -{1 \over 16})^3 \, (x-{1 \over 4})\,\, P_3 \cdot D_x^3
\,\,+x^2 \cdot (x -{1 \over 16})^{2} \,\, P_2 \cdot D_x^2 \nonumber \\
&& + x \cdot (x-{1 \over 16}) \,\, P_1 \cdot D_x\, \,+ x  \cdot  P_0
\end{eqnarray}
The polynomials $P_3, \cdots,\,  P_0$ are of degrees, respectively,
4, 6, 7 and 7 in $x$.
We have 27 coefficients (not counting the overall one) to 
reconstruct.
For easy labeling, these polynomials are denoted as ($P_3$ is the polynomial
whose roots are apparent singularities)
\begin{eqnarray}
P_3 =\,  x^4+\sum_{k=0}^{3} a_k\, x^k, \,\, P_2 =\,  \sum_{k=0}^{6} b_k\, x^k,
\,\, P_1 = \, \sum_{k=0}^{7} c_k\, x^k, \,\, P_0 = \, \sum_{k=0}^{7} d_k\, x^k.
 \nonumber
\end{eqnarray}
The indicial exponents obtained with both ODEs (with the two primes 32749 and 32719)
are
\begin{eqnarray}
x=0, &\quad& \rho = \, -2, 0, 2,  \nonumber \\
x=\infty, &\quad& \rho = 1, 2, 5/2,   \nonumber \\
x=1/16, &\quad& \rho = \, -15/4, -13/4, -1,   \nonumber \\
x=1/4, &\quad& \rho =\,  0, 1, 7/2,   \nonumber \\
P_3(\alpha)=0, &\quad& \rho = \, 0, 1, 3. \nonumber
\end{eqnarray}
By demanding that the ODE corresponding to the almost generic $\tilde{L}_3$
gives the above indicial exponents,   leads to some conditions on the
unknown coefficients $a_k, b_k, c_k$ and $d_k$.
The order of the ODE being 3, we  obtain for each singularity a maximum of three
conditions. This is a maximum, because some exponents are by construction
automatically satisfied. For instance, at $x=\, 1/4$, we  obtain only one
condition related to the exponent $\rho=\, 7/2$.

At the singularity $x=\, 0$, the indicial equation of $\tilde{L}_3$  
gives $\rho=\, 0$ as a root automatically satisfied and a polynomial in $\rho^2$
depending on some of the unknown coefficients of $\tilde{L}_3$.
By requiring $\rho=\, -2$ and $\rho=\, 2$ as roots of this polynomial, we obtain
\begin{eqnarray}
b_0 \, = \,\,  { 3 \over 64}\, a_0, \qquad c_0 \, =\,\,   { 3 \over 1024}\, a_0
\end{eqnarray}
With these values   assigned, we require that $\rho=1, 2, 5/2$ be roots of
the indicial equation at the singularity $x=\infty$. One then gets
\begin{eqnarray}
b_6\,  =\, \,  { 17 \over 2}, \qquad c_7\,  =\, \,  16, \qquad d_7\,  =\,\,   5.
\end{eqnarray}
Similarly, the indicial equations evaluated at the local exponents for
the singularities $x=1/16$ and $x=1/4$ give four equations, fixing (e.g.) 
the coefficients $b_4$, $b_5$, $c_6$ and $d_6$ in terms of   other
coefficients.

Next we turn   to the apparent singularies. These are given by the roots of $P_3$.
Calling $\alpha$ a root of $P_3$ (with unknown $a_k$), the indicial equation
appears with $\rho=\, 0$ and $\rho=\, 1$ as automatically satisfied roots.
Requiring $\rho=\, 3$ as root of the indicial equation, gives a polynomial
in $\alpha$ of degree three. Zeroing each term gives 22 solutions.
Discarding all the solutions where a coefficient from $\tilde{L}_3$ is zero,
one is left with five solutions. From these solutions, there is only one solution
which is acceptable, because it matches with the actual values of the coefficients
known in prime. This fixes three coefficients in terms of the others.

At this point, we have fixed 12 coefficients among the 27  using
only the knowledge about the local exponents.
The condition on the local exponents at the apparent singularities is only
necessary, the sufficient condition is the absence of logarithmic solutions
around the singularity $x=\, \alpha$.

The conditions on the non-occurrence of logarithmic solutions at the apparent
singularities can be imposed either by requiring the conditions of
eq.~(A.8) in~\cite{bo-gu-ha-je-ma-ni-ze-08} to be fulfilled or equivalently
by zeroing the coefficients in front of the log's in the formal solutions
of $\tilde{L}_3$ at $\alpha$.
With a generic apparent polynomial, the calculations can be cumbersome.
So let us fix some coefficients.

One finds that the ratio $-2\, a_1/a_0$ appears with both primes 32749 and
32719 as the number 63. Also for both primes one obtains $4\,a_2/a_0 =\, 839$,
$-8\,a_3/a_0 =\, 3607$ and $2^{14}\, d_0/a_0 =\, 147$. 

Furthermore, one may compute the (analytical at $x=\, 0$) series at both
primes in the hope that some coefficients will be ``simple" enough to be
recognized.
The series with the prime 32749 gives
\begin{eqnarray}
x^2 + 48\,x^3 + 1527\, x^4 + 7541 \, x^5 + 3199 \, x^6 + \cdots
\end{eqnarray}
while with the prime 32719, it reads
\begin{eqnarray}
x^2 + 48\,x^3 + 1527\, x^4 + 7571 \, x^5 + 4069 \, x^6 + \cdots
\end{eqnarray}
We note that the same values occur at  orders 3 and 4. These
 numbers are therefore likely to be exact.
Also the difference between the values at order 5 is a multiple
of the difference $32749\, -32719$, and similarly at  order 6.
It is easy to ``guess" these values as respectively, 48, 1527, 40290 and 952920.
Comparing with the series solution of $\tilde{L}_3$ fixes four coefficients.

We have then twelve coefficients fixed exactly and nine coefficients fixed
by reconstruction.
The formal solutions of $\tilde{L}_3$ at the apparent singularity $\alpha$
give two logarithmic solutions, with leading term, each
\begin{eqnarray}
C \,\, \alpha^k \,\, (x-\alpha)^3 \, \ln(x-\alpha), \quad \quad  \quad k=\, 0, \cdots, \, 3 
\end{eqnarray}
where $C$ depends on the remaining non fixed coefficients of $\tilde{L}_3$.
We have then eight (non-linear) equations for six unknowns to solve.
This can be done by rational reconstruction and check.

\section{The differential operator $L_4$ in exact arithmetics}
\label{appL4}

The degree $n$ polynomials $P_j^{(n)}(x)$ occurring in the differential operator $L_4$ read:
\begin{eqnarray}
\fl P_4^{(26)} =
28000-7854000\,x+873083400\,{x}^{2}-54037012120\,{x}^{3}+2099285510560\,{x}^{4} \nonumber \\
\fl \quad -52582954690298\,{x}^{5}+766418384173454\,{x}^{6}-1305110830870633\,{x}^{7} \nonumber \\
\fl \quad   -251264549473230968\,{x}^{8}+7727889974481947660\,{x}^{9} \nonumber \\
\fl \quad   -148605250583921845896\,{x}^{10}+2252938824290334087840\,{x}^{11} \nonumber \\
\fl \quad   -29645475671183771992224\,{x}^{12}+354446803792968575565792\,{x}^{13} \nonumber \\
\fl \quad   -3850023960384577768909952\,{x}^{14}+36761552740911534545901568\,{x}^{15} \nonumber \\
\fl \quad   -296338746597146803591135232\,{x}^{16}+1953967934450852091348254720\,{x}^{17} \nonumber \\
\fl \quad   -10332892566359614848157876224\,{x}^{18}+43345424617004971574289235968\,{x}^{19} \nonumber \\
\fl \quad   -142807225508285034141616963584\,{x}^{20}+359505820412663945726355570688\,{x}^{21} \nonumber \\
\fl \quad   -636026962079787427490890252288\,{x}^{22}+616797192523902897669611192320\,{x}^{23} \nonumber \\
\fl \quad   +45081769872830521912080728064\,{x}^{24}-724445324775545659452335063040\,{x}^{25} \nonumber \\
\fl \quad   +521686412421099571093753036800\,{x}^{26}, \nonumber 
\end{eqnarray}
\begin{eqnarray}
\fl P_3^{(33)} =
-4480000+1569568000\,x-238072889600\,{x}^{2}+21281848471520\,{x}^{3} \nonumber \\
\fl \quad  -1268595101537120\,{x}^{4}+53555230610961720\,{x}^{5}-
1640958998875092768\,{x}^{6} \nonumber \\
\fl \quad   +36032181180727162732\,{x}^{7}-511428562675996247108\,{x}^{8} \nonumber \\
\fl \quad   +1919885419260765103140\,{x}^{9}+129005457127386313373184\,{x}^{10} \nonumber \\
\fl \quad   -4541113747259527374959592\,{x}^{11}+96035689755227434986877112\,{x}^{12} \nonumber \\
\fl \quad   -1580421468708164786235613784\,{x}^{13}+22087897691588508601005658336\,{x}^{14} \nonumber \\
\fl \quad   -274269909442085751262554453856\,{x}^{15}+3087338965228238905750107987648\,{x}^{16} \nonumber \\
\fl \quad   -31474922613166692487806647824256\,{x}^{17}   \nonumber \\
\fl \quad   +286292076483602608978320943481344\,{x}^{18} \nonumber \\
\fl \quad   -2277952733740370146287983798312960\,{x}^{19} \nonumber \\
\fl \quad   +15571420858521621122719931928608768\,{x}^{20} \nonumber \\
\fl \quad   -90147310596750652057735905075527680\,{x}^{21}  \nonumber \\
\fl \quad   +437037767842717994841190340774330368\,{x}^{22} \nonumber \\
\fl \quad   -1755686044559298411692425577783885824\,{x}^{23} \nonumber \\
\fl \quad   +5764646607249819312839063743970148352\,{x}^{24} \nonumber \\
\fl \quad   -15099644256008129321411266837095645184\,{x}^{25} \nonumber \\
\fl \quad   +29988658044590195137583404663925899264\,{x}^{26} \nonumber \\
\fl \quad   -39745090934862435542362760545321353216\,{x}^{27}  \nonumber \\
\fl \quad   +19398167217699147074111209484113149952\,{x}^{28} \nonumber \\
\fl \quad   +40447257076217292533523320942836580352\,{x}^{29}  \nonumber \\
\fl \quad   -84060042791775646091063152898350252032\,{x}^{30} \nonumber \\
\fl \quad   +40724840987587942318458896159738953728\,{x}^{31}  \nonumber \\
\fl \quad   +34088801304111660197683822288919592960\,{x}^{32} \nonumber \\
\fl \quad   -34873025538917765121024203000119296000\,{x}^{33}, \nonumber 
\end{eqnarray}
\begin{eqnarray}
\fl P_3^{(33)} =
-4480000+1569568000\,x-238072889600\,{x}^{2}+21281848471520\,{x}^{3} \nonumber \\
\fl \quad    -1268595101537120\,{x}^{4}+53555230610961720\,{x}^{5}-
1640958998875092768\,{x}^{6} \nonumber \\
\fl \quad    +36032181180727162732\,{x}^{7}-511428562675996247108\,{x}^{8} \nonumber \\
\fl \quad     +1919885419260765103140\,{x}^{9}+129005457127386313373184\,{x}^{10} \nonumber \\
\fl \quad     -4541113747259527374959592\,{x}^{11}+96035689755227434986877112\,{x}^{12} \nonumber \\
\fl \quad     -1580421468708164786235613784\,{x}^{13}+22087897691588508601005658336\,{x}^{14} \nonumber \\
\fl \quad     -274269909442085751262554453856\,{x}^{15}+3087338965228238905750107987648\,{x}^{16} \nonumber \\
\fl \quad     -31474922613166692487806647824256\,{x}^{17} \nonumber \\
\fl \quad    +286292076483602608978320943481344\,{x}^{18} \nonumber \\
\fl \quad     -2277952733740370146287983798312960\,{x}^{19}      \nonumber \\
\fl \quad    +15571420858521621122719931928608768\,{x}^{20} \nonumber \\
\fl \quad     -90147310596750652057735905075527680\,{x}^{21}     \nonumber \\
\fl \quad    +437037767842717994841190340774330368\,{x}^{22} \nonumber \\
\fl \quad     -1755686044559298411692425577783885824\,{x}^{23}   \nonumber \\
\fl \quad    +5764646607249819312839063743970148352\,{x}^{24} \nonumber \\
\fl \quad     -15099644256008129321411266837095645184\,{x}^{25}  \nonumber \\
\fl \quad    +29988658044590195137583404663925899264\,{x}^{26} \nonumber \\
\fl \quad     -39745090934862435542362760545321353216\,{x}^{27}  \nonumber \\
\fl \quad    +19398167217699147074111209484113149952\,{x}^{28} \nonumber \\
\fl \quad     +40447257076217292533523320942836580352\,{x}^{29}  \nonumber \\
\fl \quad    -84060042791775646091063152898350252032\,{x}^{30} \nonumber \\
\fl \quad     +40724840987587942318458896159738953728\,{x}^{31}  \nonumber \\
\fl \quad    +34088801304111660197683822288919592960\,{x}^{32} \nonumber \\
\fl \quad     -34873025538917765121024203000119296000\,{x}^{33}, \nonumber 
\end{eqnarray}
\begin{eqnarray}
\fl P_2^{(38)} =
202496000-84671104000\,x+15961404659200\,{x}^{2}  -1817819283938560\,{x}^{3} \nonumber \\
\fl \quad     +141042261097575040\,{x}^{4}-7945786419559994432\,{x}^{5}+336970482890735391136\,{x}^{6} \nonumber \\
\fl \quad     -10948102706558839518064\,{x}^{7}+272101251799491505044720\,{x}^{8}                      \nonumber \\
\fl \quad     -4990110947182458236154960\,{x}^{9}+57747165172968723279034760\,{x}^{10}                 \nonumber \\
\fl \quad     +4251375690841730042108460\,{x}^{11}-19664102412813111595220034000\,{x}^{12}             \nonumber \\
\fl \quad     +586539601535060491103255831780\,{x}^{13}-11648280832868820874871506994648\,{x}^{14}                 \nonumber \\
\fl \quad     +185168754164459407231412940918408\,{x}^{15}     \nonumber \\
\fl \quad    -2524027149739792644483439537740736\,{x}^{16}            \nonumber \\
\fl \quad     +30612264160202676427790224166656736\,{x}^{17}   \nonumber \\
\fl \quad    -336756368430758251349398256374549440\,{x}^{18}        \nonumber \\
\fl \quad     +3374928905004383352843307682288939648\,{x}^{19}   \nonumber \\
\fl \quad    -30594829577694461795851875047251759104\,{x}^{20}     \nonumber \\
\fl \quad     +247550135999641906395555078053550042624\,{x}^{21}    \nonumber \\
\fl \quad    -1761694860791556801623390940580862476288\,{x}^{22}       \nonumber \\
\fl \quad     +10880585300165439414813579169355207311360\,{x}^{23}   \nonumber \\
\fl \quad    -57651463831886251900194559835893548711936\,{x}^{24}    \nonumber \\
\fl \quad     +259270510361927197193957311877476593434624\,{x}^{25}   \nonumber \\
\fl \quad    -977978052427489585499761822245900827754496\,{x}^{26}  \nonumber \\
\fl \quad     +3043305515555663644471442318841392857612288\,{x}^{27}   \nonumber \\
\fl \quad    -7593091629989468917294503828609326603304960\,{x}^{28} \nonumber \\
\fl \quad     +14317720902933442365662690637880059263713280\,{x}^{29}   \nonumber \\
\fl \quad    -17337418172194871339769688830180251691646976\,{x}^{30} \nonumber \\
\fl \quad     +3546879809404692840748046019057281136590848\,{x}^{31}  \nonumber \\
\fl \quad     +32904223733304447370184725984806679848419328\,{x}^{32}  \nonumber \\
\fl \quad     -61070255095717193234874579385327453575577600\,{x}^{33} \nonumber \\
\fl \quad     +27287160011587832026533318214423160423448576\,{x}^{34} \nonumber \\
\fl \quad     +47538188516382446352727572349627507901726720\,{x}^{35} \nonumber \\
\fl \quad     -56445574686008125172119780480189438504206336\,{x}^{36} \nonumber \\
\fl \quad     -4318904703692797702055795738669690860339200\,{x}^{37}  \nonumber \\
\fl \quad     +23615008551819589708322104774634383815475200\,{x}^{38}, \nonumber 
\end{eqnarray}
\begin{eqnarray}
\fl P_1^{(43)} =
-2508800000+1313872896000\,x-313495056179200\,{x}^{2}   \nonumber \\
\fl \quad    +45402581315051520\,{x}^{3} -4502030899704432640\,{x}^{4}+326696241278915100672\,{x}^{5} \nonumber \\
\fl \quad    -18076764858722283537408\,{x}^{6}  +782686127310817603163904\,{x}^{7} \nonumber \\
\fl \quad    -26913654199485748976447296\,{x}^{8} +737895426074343351817982240\,{x}^{9}         \nonumber \\
\fl \quad   -15941906513987915790530627104\,{x}^{10} +258773331815879690900773968400\,{x}^{11}    \nonumber \\
\fl \quad    -2607962306360230233373492782176\,{x}^{12}  -5952736815704779243433578988240\,{x}^{13}       \nonumber \\
\fl \quad   +986994067078072761785220512495568\,{x}^{14}     \nonumber \\
\fl \quad  -27146844884553280870530528088810192\,{x}^{15}           \nonumber \\
\fl \quad  +517853080131584647940304813906843912\,{x}^{16}              \nonumber \\
\fl \quad     -8012481021063135055260920360860291792\,{x}^{17}    \nonumber \\
\fl \quad  +106770798207835443855237151398845884336\,{x}^{18}         \nonumber \\
\fl \quad     -1266450623115899739824560105189293118336\,{x}^{19}    \nonumber \\
\fl \quad  +13633188914202542686825030295715897195712\,{x}^{20}           \nonumber \\
\fl \quad     -134327854309114583892390549684219213327360\,{x}^{21}    \nonumber \\
\fl \quad  +1210064015789594600623288132568732268617728\,{x}^{22}       \nonumber \\
\fl \quad     -9885950925613173310943030286291198265745664\,{x}^{23}     \nonumber \\
\fl \quad  +72409258384998425181940033322470418064579584\,{x}^{24}     \nonumber \\
\fl \quad     -469743224562744760675515167582702668512534528\,{x}^{25}        \nonumber \\
\fl \quad     +2668377435339727605940145954082991900173762560\,{x}^{26}       \nonumber \\
\fl \quad     -13130369247854115699188867418934857934469332992\,{x}^{27}      \nonumber \\
\fl \quad     +55351059606988527054614355506855140926785847296\,{x}^{28}    \nonumber \\
\fl \quad     -197223974876465329508996857363261585516738379776\,{x}^{29}     \nonumber \\
\fl \quad     +582489670248346892198679343375535443002292961280\,{x}^{30}  \nonumber \\
\fl \quad     -1378041300571967278991550115889940047326999478272\,{x}^{31}    \nonumber \\
\fl \quad     +2424920581794299143009574014342980105306252509184\,{x}^{32}  \nonumber \\
\fl \quad     -2497995923785565959357957923036374193256014020608\,{x}^{33}    \nonumber \\
\fl \quad     -915221239768968177447587513000776938544384966656\,{x}^{34}   \nonumber \\
\fl \quad     +8834509277491743877951520843172281230968074797056\,{x}^{35}    \nonumber \\
\fl \quad     -14739433233907061551935598195219565259254451404800\,{x}^{36} \nonumber \\
\fl \quad     +5672247850181350880926829983674652540556997033984\,{x}^{37}  \nonumber \\
\fl \quad     +16865676565374917674763783610716646680263347142656\,{x}^{38} \nonumber \\
\fl \quad     -23057876384647717319687179507181219615528382889984\,{x}^{39} \nonumber \\
\fl \quad     -1650497416603706423024253737913692788809736912896\,{x}^{40} \nonumber \\
\fl \quad     +17950328578610802277327697770535083294270546771968\,{x}^{41} \nonumber \\
\fl \quad     -3724970001243182786619005376755715492471782768640\,{x}^{42} \nonumber \\
\fl \quad     -5981413341400069058756778898532874294556360704000\,{x}^{43}, \nonumber 
\end{eqnarray}
\begin{eqnarray}
\fl P_0^{(47)}/16 =
 58841859686400\,x-123282432000-13099552866570240\,{x}^{2} \nonumber \\
\fl \quad     +1817269523720161280\,{x}^{3}-176880012691621796864\,{x}^{4}                         \nonumber \\
\fl \quad     +12880441893460632329216\,{x}^{5}-729910851392566766105088\,{x}^{6}                  \nonumber \\
\fl \quad     +33014141392329879832166784\,{x}^{7}-1210942171584302533599014752\,{x}^{8}           \nonumber \\
\fl \quad     +36308844474092544015578885632\,{x}^{9}-889129088058672919373638221264\,{x}^{10}                 \nonumber \\
\fl \quad     +17508381271590013090109310169040\,{x}^{11}      \nonumber \\
\fl \quad    -263494886656617518206756373588932\,{x}^{12}         \nonumber \\
\fl \quad   +2493591715504400008185935972185648\,{x}^{13}            \nonumber \\
\fl \quad   +5772652777357046820837948335210000\,{x}^{14}       \nonumber \\
\fl \quad     -880112646375062548999453842320020740\,{x}^{15}       \nonumber \\
\fl \quad   +23493494316713860255651067234081149257\,{x}^{16}              \nonumber \\
\fl \quad     -438058417861614862884693737035489286345\,{x}^{17}      \nonumber \\
\fl \quad   +6643943767863126335261566491851505292189\,{x}^{18}         \nonumber \\
\fl \quad     -86756699901268061114746560625886582904365\,{x}^{19}      \nonumber \\
\fl \quad   +1006186625234680761751520210312145980149549\,{x}^{20}    \nonumber \\
\fl \quad     -10573522222931154420271493264607253396390520\,{x}^{21}      \nonumber \\
\fl \quad   +101894357518227690884588911318694326634020120\,{x}^{22}         \nonumber \\
\fl \quad     -904489874897014837177389619617321458811380360\,{x}^{23} \nonumber \\
\fl \quad     +7377025197157259422622822297421365236335307120\,{x}^{24}       \nonumber \\
\fl \quad     -54857750379533672661182684179932897350723993600\,{x}^{25}   \nonumber \\
\fl \quad     +368077157510764846299472690339090755869412496960\,{x}^{26}   \nonumber \\
\fl \quad     -2203473576836831766402446311571835988588370992640\,{x}^{27}  \nonumber \\
\fl \quad     +11638948368194240385082022186232592838207372154880\,{x}^{28} \nonumber \\
\fl \quad     -53640316561668843524196008022033100643589470191616\,{x}^{29}     \nonumber \\
\fl \quad     +213035315257870225008043406258186703365521254907904\,{x}^{30}    \nonumber \\
\fl \quad     -717584853510007605413068883853037631249102250442752\,{x}^{31}    \nonumber \\
\fl \quad     +2000779126900084461641442809746125018650394950107136\,{x}^{32}   \nonumber \\
\fl \quad     -4414463297786097513664235192893813927566161255333888\,{x}^{33}   \nonumber \\
\fl \quad     +6904891435787610921130882736916279097844736823656448\,{x}^{34}   \nonumber \\
\fl \quad     -4551724050684467601081502988404586537388373763424256\,{x}^{35}   \nonumber \\
\fl \quad     -11571837065995769727688883612933393577503693010370560\,{x}^{36}  \nonumber \\
\fl \quad     +43604071314966497936511910817544815142484611319201792\,{x}^{37}  \nonumber \\
\fl \quad     -64046695475293378492360343847354456353947960484036608\,{x}^{38}  \nonumber \\
\fl \quad     +17229520899952062417015850756255391466062797246300160\,{x}^{39}  \nonumber \\
\fl \quad     +98913305027317465024954824787190137389180923821424640\,{x}^{40}  \nonumber \\
\fl \quad     -145693357979556257119098588624331861246512084740472832\,{x}^{41} \nonumber \\
\fl \quad     -1743763200037842518500493452602647084799741447372800\,{x}^{42}   \nonumber \\
\fl \quad     +159968299464829816606333313819738117801053481636724736\,{x}^{43} \nonumber \\
\fl \quad     -68453133710464189864730237770717937227743579749744640\,{x}^{44}  \nonumber \\
\fl \quad     -84974525390992986946108353023934616304288806232653824\,{x}^{45}  \nonumber \\
\fl \quad     +41407097440632033071894561954752886956699467613470720\,{x}^{46}  \nonumber \\
\fl \quad     +28698609854675644415679733396189051258415886630912000\,{x}^{47}.  
 \nonumber
\end{eqnarray}

\end{document}